\documentclass[preprint]{aastex701}
\usepackage{graphicx} 
\usepackage{float}

\begin{document}

\title{The long-term accretion luminosity of V4641~Sgr through binary evolution simulations: implications for its ultrahigh-energy gamma-ray emission}

\author[orcid=0000-0003-1576-0961]{Ruo-Yu Liu}
\affiliation{School of Astronomy and Space Science, Nanjing University, Nanjing 210023, China}
\affiliation{Key laboratory of Modern Astronomy and Astrophysics(Nanjing University),Ministry of Education,Nanjing 210023, People's Republic of China}
\affiliation{Tianfu Cosmic Ray Research Center,
 Chengdu 610000, Sichuan, China}
\email{ryliu@nju.edu.cn}

\author[orcid=0000-0003-2506-6906]{Yong Shao} 
\affiliation{School of Astronomy and Space Science, Nanjing University, Nanjing 210023, China}
\affiliation{Key laboratory of Modern Astronomy and Astrophysics(Nanjing University),Ministry of Education,Nanjing 210023, People's Republic of China}
\email{shaoyong@nju.edu.cn}

\author[orcid=0009-0009-4482-6350]{Yu-Dong Nie} 
\affiliation{School of Astronomy and Space Science, Nanjing University, Nanjing 210023, China}
\affiliation{Key laboratory of Modern Astronomy and Astrophysics(Nanjing University),Ministry of Education,Nanjing 210023, People's Republic of China}
\email{1452531593@qq.com}

\correspondingauthor{Yong Shao; Ruo-Yu Liu}
\email{shaoyong@nju.edu.cn; ryliu@nju.edu.cn}

\begin{abstract}
Recent observations by LHAASO and HAWC have revealed extended ultrahigh-energy (UHE; $E>100$~TeV) gamma-ray emission associated with the black-hole X-ray binary (BHXRB) V4641~Sgr, with a spectrum extending up to $\sim0.8$~PeV. Interpreting this emission requires a very high time-averaged non-thermal particle power, significantly exceeding {the long-term observed X-ray luminosity which is commonly used as a proxy for the accretion power}, leading to an apparent ``energy crisis''. To address this, we perform detailed binary-evolution simulations with \textit{MESA}, constrained by the known system parameters inferred from observation. Across an extensive evolutionary grid, all tracks that match the current system parameters pass through a long-lasting, slow mass-transfer phase, with a time-averaged intrinsic X-ray luminosity of over evolutionary timescales of order $L_X\sim10^{38}$~erg~s$^{-1}$, far above the observed luminosity average over the last few decades. This is consistent with earlier suggestions of an extended obscuring/reprocessing envelope or outflow in V4641~Sgr. The inferred intrinsic accretion power can then readily supply the energy required to explain the UHE emission under the leptonic model, and is also marginally consistent with the requirement from the hadronic model, resolving the energy crisis. This supports V4641~Sgr as a Galactic PeV particle accelerator.
\end{abstract}

\section{Introduction}
\label{sec:intro}

V4641~Sgr is a black hole X-ray binary (BHXRB) system consisting of a $6.4\pm0.6\,M_\odot$ black hole (BH) and a $2.9\pm0.4\,M_{\odot}$ B9III donor star of an effective temperature of $T_{\rm eff}\approx 10500\,$K \citep{Orosz2001, Sadakane2006, MacDonald2014}. Located at a distance of $6.2 \pm 0.7$~kpc \citep{MacDonald2014}, it has an orbital period of $\sim 2.82$~days \citep{Orosz2001}. Its identification is via an extremely luminous and rapid outburst in 1999 with super-Eddington luminosity lasting about 2 hours \citep{Smith1999, in'tZand2000, Wijnands2000}. Multi-wavelength studies, including radio observations revealing apparent superluminal motion, suggest the presence of relativistic jets \citep{Hjellming2000}, classifying it as a microquasar. Since its discovery, V4641~Sgr has exhibited recurrent outbursts with a median recurrence time of  a few years \citep{Tetarenko2016}. However,  none of the subsequent outbursts (which had X-ray luminosities of $L_{\rm X} \sim 10^{-2}-10^{-4}\ L_{\rm Edd}$) matched the extreme luminosity of the 1999 event \citep{Bailyn2003, Yamaoka2010, Tachibana2014, Bahramian2015, Pahari2015, Negoro2018, Shaw2022, Parra2025}.
In addition to outbursts, V4641~Sgr spends long periods of time in quiescent state, characterized by low X-ray luminosities $L_{\rm X}<10^{-4}\,L_{\rm Edd}$, which could be even down to an extremely low level of $10^{-8}L_{\rm Edd}$ \citep{Tomsick2003}. The long-term average luminosity is difficult to determine because continuous X-ray coverage is unavailable, but it is likely much lower than $10^{-4}L_{\rm Edd}$, or $\sim 10^{35}\,\mathrm{erg\,s^{-1}}$, and is not represented by the luminosity measured during outbursts.

Recently, extended ultrahigh-energy (UHE; photon energy $E>100\,$TeV) gamma-ray emission associated to V4641~Sgr \citep{LHAASO2025_microquasar, HAWC2024} has been discovered by the Large High Altitude Air Shower Observatory (LHAASO) and the High Altitude Water Cherenkov Observatory (HAWC), and extended TeV emission was recently reported by the High Energy Stereoscopic System (H.E.S.S.) \citep{HESS2025}.  The gamma-ray source exhibits an elongated morphology with a projected size of $\sim 100$~pc, suggesting that a huge amount of energy in non-thermal particles has been injected from the BH-jet system. The gamma-ray spectrum peaks around 30\,TeV, extending up to $\sim 0.8$~PeV \citep{LHAASO2025_microquasar}, with a total gamma-ray luminosity of $1.3\times 10^{34}\,$erg/s in the UHE band, exceeding the observed luminosities of other PeV gamma-ray emitters, such as some young PWNe \citep{LHAASO2021_Crab, LHAASO2026_J1849}, Cygnus Bubble \citep{LHAASO2024_CB}, and Cygnus~X-3 \citep{LHAASO2025_X3}, by about one order of magnitude.  V4641~Sgr seems to be the most powerful Galactic PeVatrons known.

Interpreting the UHE gamma-ray emission poses significant challenges. The radiative timescale associated with the UHE emission is at least tens of thousands of years \citep{Wan2026}. Therefore, the formation of the extended UHE structure around V4641~Sgr is tied to the long-term, time-integrated energy output of the binary system. In accreting black-hole binaries, the observed X-ray luminosity is often used as an empirical proxy for the instantaneous accretion power. If such a proxy is extrapolated to the long-term particle-injection budget relevant for the extended UHE emission, V4641~Sgr appears energetically challenging whether in leptonic or hadronic scenarios, leading to the ``energy crisis''. Hadronic models, which attribute the emission to proton-proton collisions \citep{Romero2003, Bosch-Ramon2005, Neronov2025}, suffer very low radiation efficiency because the density of the surrounding medium is constrained to be $n_{\rm H}<0.2\,\rm  cm^{-3}$ \citep{HESS2025}, requiring a continuous injection of PeV protons with a luminosity of $\sim 5\times 10^{38}(n_{\rm H}/0.2\,\rm cm^{-3})^{-1}\,$erg/s over tens of kiloyears \citep{Wan2026, HESS2025}. This exceeds the time-averaged X-ray luminosity by orders of magnitude. Leptonic models, in which the gamma-rays originate from inverse Compton scattering of relativistic electrons, may be an alternative. The energy requirement in the leptonic model is significantly reduced compared to the hadronic model because of relatively high radiation efficiency of electrons. However, to overcome the Klein-Nishina effect and explain the elongated morphology of the gamma-ray source, \citet{Wan2026} suggested shear particle acceleration mechanism \citep{Berezhko1981, Earl1988, Rieger2006} along a hidden jet with velocity of $0.6-0.7\,c$. It leads to a kinetic power in relativistic electrons of $\sim 10^{37}\,$erg/s, still surpassing the observed long-term X-ray luminosity of V4641~Sgr.

A potential solution to this dilemma may lie in the specific viewing geometry and the consequent obscuration of the central engine, implied by the high inclination of the orbital plane ($i_{\mathrm{orb}} \approx 72^\circ$; \citealt{Orosz2001, MacDonald2014}). Indeed, \citet{Koljonen2020} demonstrated that the X-ray spectra of V4641~Sgr and other high-inclination analogs (e.g., V404~Cyg, GRS~1915+105) can be well described by models with substantial absorption or scattering in a dense equatorial outflow or a thickened accretion disk. Besides, \citet{Shaw2022}, through high-resolution \textit{Chandra} spectroscopy during the 2020 outburst, found that the measured inner disk temperatures and observed X-ray luminosities implied an unfeasibly small inner disc radius. These studies align with earlier and later studies based on optical and X-ray data from the 1999 outburst and 2024 outburst, which suggested the presence of an extended, obscuring envelope \citep{Revnivtsev2002a, Revnivtsev2002b, Parra2025}. If the long-term average intrinsic accretion luminosity has been substantially underestimated due to obscuration especially during low-flux states, the energy budget required for UHE gamma-ray production could be accommodated.

Motivated by this possibility, we perform detailed binary evolution simulations of V4641~Sgr to derive its long-term mass transfer rate and intrinsic accretion luminosity. By adopting the observed system parameters (donor mass and temperature, orbital period, etc.) as constraints on the model, we aim to derive a more realistic estimate of the time-averaged accretion power. This will allow us to reassess whether the system can provide sufficient energy to power the observed UHE gamma-ray emission, thereby addressing the energy crisis in both leptonic and hadronic scenarios.

The rest of this paper is organized as follows. In Section~\ref{sec:result}, we describe the binary evolution model, and present the simulated mass transfer history and the estimated long-term accretion luminosity. In Section~\ref{sec:discuss}, we discuss the implications for UHE gamma-ray models and compare with multi-wavelength constraints. We conclude in Section~\ref{sec:conclusion}.

\section{Evolution of V4641~Sgr}
\label{sec:result}
To determine the current evolutionary state of a binary system with physical properties similar to V4641~Sgr, we follow the systematic evolutionary study of BHXRBs performed by \citet{Shao2020} using the stellar evolution code \textit{MESA} \citep{Paxton2011, Paxton2015, Paxton2019}. In the study, a library of over 2500 evolutionary sequences has been generated, covering a wide range of initial orbital periods ($P_{\rm orb}^{\rm i}$) from $\sim 0.25$ days to 100 days (in logarithmic steps of 0.1) and initial donor masses ($M_{\rm d}^{\rm i}$) from $\sim 0.7\,M_\odot$ to $\sim 60\,M_\odot$ (in logarithmic steps of 0.025). The initial BH masses were set to either $M_{\rm BH}^{\rm i} =7\,M_\odot $ or $11\,M_\odot $. In the present work, we extend this evolutionary grid by including two lower masses: $M_{\rm BH}^{\rm i} =4\,M_\odot $ and $5\,M_\odot $. This expanded library provides a comprehensive mapping of possible evolutionary tracks for systems such as V4641~Sgr.

In the \textit{MESA} simulations, each binary is initialized with a
zero-age main-sequence donor star and a non-rotating BH in a circular orbit. The BH is modeled as a point mass. Mass transfer via Roche lobe overflow is calculated following \citet{Ritter1988}. The accretion rate onto the BH is limited by the Eddington rate,
\begin{equation}
\dot{M}_{\rm E} = \frac{4\pi G M_{\rm BH}}{\eta\kappa c},
\end{equation}
where $G$ is the 
gravitational constant, $ \kappa $ the opacity, $ c $  the speed of light, and $ \eta $ the radiative efficiency, given by 
\begin{equation}
\eta = 1-\sqrt{1- \left( \frac{M_{\rm BH}}{3M_{\rm BH}^{\rm i} } \right)^{2}} 
\end{equation}
for $ M_{\rm BH} < \sqrt{6} M_{\rm BH}^{\rm i}$ \citep{Bardeen1970}. 
%As the BH accretes mass and angular momentum, its spin parameter $ a_{\ast} $ is updated following \citet{Thorne1974}.
The transferred matter forms an accretion disk around the BH. 
Depending on whether the accretion disk is thermally and viscously unstable \citep{Lasota2001}, 
BHXRBs are classified as transient or persistent sources. The disk state is 
determined by a critical mass-transfer rate, for which we adopt the prescription of \citet{Lasota2008} for an irradiated accretion disk. 
If the mass-transfer rate falls below this critical value, the system is considered a transient source, undergoing short outbursts separated by extended quiescent phases. During quiescence, matter accumulates in the disk and is subsequently accreted during outbursts, where the accretion rate can be significantly enhanced. The BH accretion rate is capped at the Eddington limit, and any mass 
lost from the binary is assumed to carry away the specific orbital angular momentum of the BH \citep[see also][]{Shao2020}. 

\begin{figure*}[hbtp]
\centering
\includegraphics[width=0.8\textwidth]{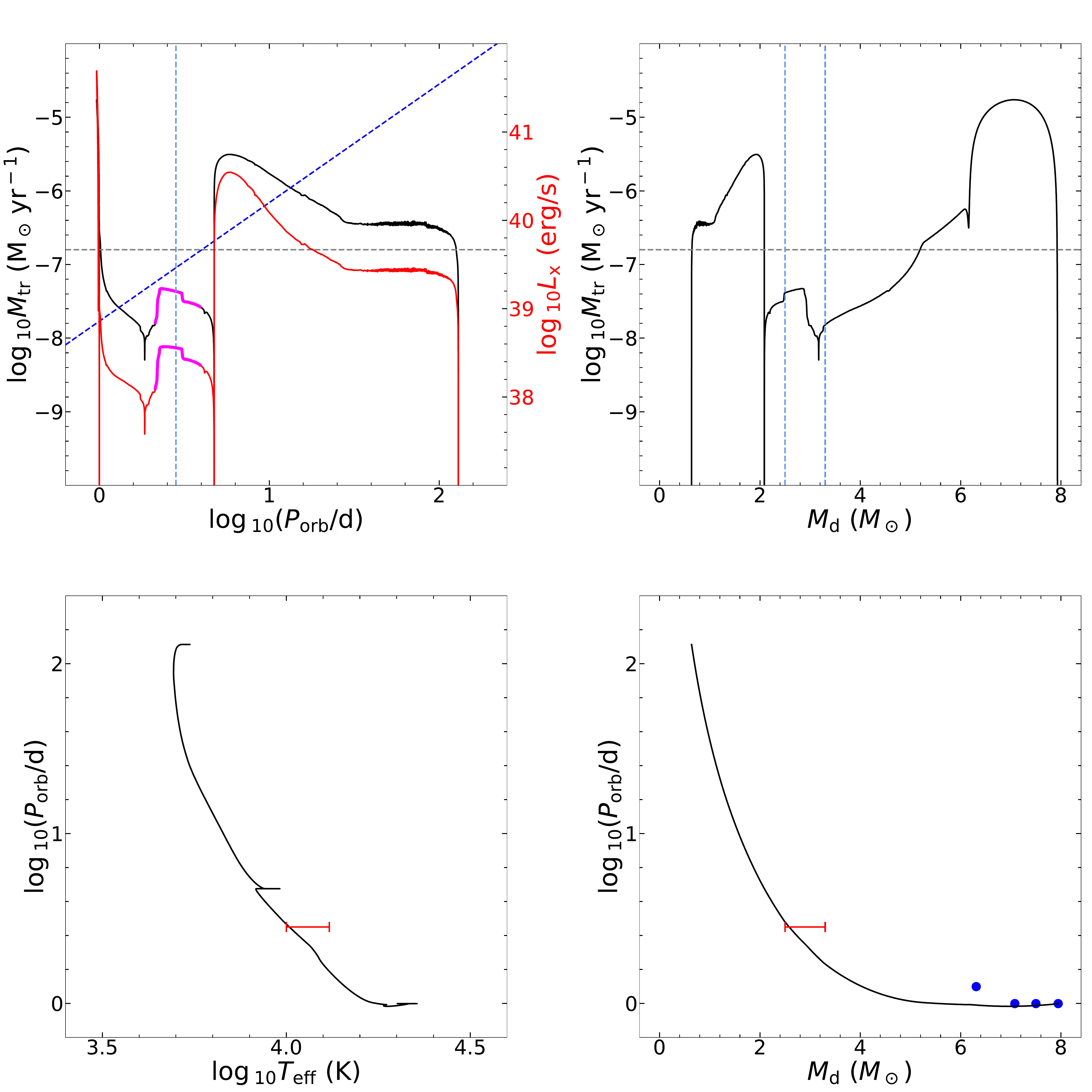}
%\linespread{0.7}
\caption{Evolutionary track for a binary system initially composed of a $5\,M_\odot$ BH and a $7.9\,M_\odot$ donor in a 1-day orbit. {\bf Upper-left panel}: Mass-transfer rate $\dot{M}_{\rm tr}$ (black curve) and X-ray luminosity $L_{\rm X}$ (red curve) as a function of orbital period $P_{\rm orb}$. The vertical dashed line marks the observed orbital period of V4641~Sgr ($P_{\rm orb}=2.82$ days). The thick magenta curves denote a 20~Myr time interval centered on this orbital period. The horizontal dashed line indicates the Eddington accretion rate $\dot{M}_{\rm Edd}$ for the BH. The slanted dashed line represents the critical mass-transfer rates below which the accretion disk becomes unstable and the binary appears as a transient source \citep{Lasota2008}. {\bf Upper-right panel}: Mass-transfer rate as a function of donor mass $M_{\rm d}$. The observed donor mass range for V4641~Sgr ($2.5-3.3\,M_\odot$) is delineated by two vertical
dashed lines. The horizontal dashed line has the same meaning as that in the upper-left panel.} {\bf Lower-left panel}: Orbital period versus donor effective temperature $T_{\rm eff}$. The red symbol indicates the observed position of V4641~Sgr. At this position, the modeled donor has $\log\,g\sim3.4$, consistent with spectroscopic constraints  \citep{Orosz2001}. {\bf Lower-right panel}: Orbital period versus donor mass. The four blue dots represent initial binary configurations (selected from our evolutionary library) that evolve into systems resembling V4641~Sgr.  \label{V4641sgrf1}
\end{figure*}

For sub-Eddington accretion, the X-ray luminosity is related to the mass transfer rate $\dot{M}_{\rm tr}$ via 
\begin{equation}
L_{\rm X}=\eta\dot{M}_{\rm tr}c^{2}.
\end{equation}
When $\dot{M}_{\rm tr} $ exceeds the Eddington 
rate $ \dot{M}_{\rm E} $ during the evolution,
the accretion disk becomes geometrically thick, strongly influencing the emergent X-ray luminosity. 
In this regime, the luminosity can be estimated as 
\begin{equation}
L_{\rm X} \simeq \frac{L_{\rm E}}{b}\left[1+\ln \left(
\frac{\dot{M}_{\rm tr}}{\dot{M}_{\rm E}} \right) \right],
\end{equation}
where $ b $ is the beaming factor. Following
\citet{King2009}, we adopt the approximate expression
 \begin{equation}
b \simeq \frac{73}{\dot{m}^2}, 
\end{equation}
with $ \dot{m} = \dot{M}_{\rm tr} / \dot{M}_{\rm E}$. This formula applies when $  \dot{m} \gtrsim 8.5$, otherwise,  
beaming is negligible (i.e., $ b = 1$). All simulations are terminated when the donor becomes degenerate or dynamically unstable mass transfer ensues \cite[see][for details]{Shao2020}.

From the library of evolutionary tracks, we select binaries resembling V4641~Sgr by imposing constraints on orbital period ($P_{\rm orb} = 2.81-2.83$ days), donor mass ($M_{\rm d} = 2.5-3.3\,M_\odot$), and effective temperature ($\log_{10}\, (T_{\rm eff}/\rm K) = 4.0-4.1$) \citep{MacDonald2014}. We identify {seven systems from the full library that reproduce the observed properties of V4641~Sgr at some stage of their evolution. For all of these matching systems, the initial donor masses range from $\sim 6\,M_\odot$ to $\sim 8\,M_\odot$ and initial orbital periods lie between $\sim 1$ day and 2 days. We then examine their mass-transfer rates and intrinsic X-ray luminosities at the epochs when they match the observed properties of V4641~Sgr's.}

Figure~\ref{V4641sgrf1} shows the evolutionary track of a binary initially consisting of a $5\,M_{\odot}$ BH and a $7.9\,M_{\odot}$ donor star in a 1-day orbit. Roche lobe overflow begins when the donor is still on the main sequence. Due to the relatively large initial mass ratio, the system first experiences rapid mass transfer on the donor’s thermal timescale, with $\dot{M}_{\rm tr}\sim10^{-6}-10^{-5}\,M_{\odot}\,\rm yr^{-1}$. Such a high rate could render the binary an ultraluminous X-ray source \citep{King2009,Podsiadlowski2003}. During this phase, the donor loses $\sim 2\,M_\odot$ of its hydrogen envelope. Subsequently, the donor remains near thermal equilibrium, and mass transfer becomes entirely driven by its nuclear evolution. The mass-transfer rate declines from $\sim10^{-6}\,M_{\odot}\,\rm yr^{-1}$ to $\sim10^{-8}\,M_{\odot}\,\rm yr^{-1}$, transferring about $ 3\,M_\odot$ from the donor to the BH over this phase and increasing the BH mass to $\gtrsim7\,M_\odot$. V4641~Sgr is likely undergoing such a slow mass-transfer phase. As the donor approaches the end of the main sequence, it becomes detached, temporarily halting mass transfer. Later, as the donor ascends the giant branch, the mass-transfer rate rises sharply to $\sim3\times 10^{-6}\,M_{\odot}\,\rm yr^{-1}$ and then decreases to $\sim3\times 10^{-7}\,M_{\odot}\,\rm yr^{-1}$. Over $1\,M_\odot$ of the donor's envelope is transferred during this final mass-transfer phase. Throughout this evolution, the orbital period increases from 1 day to $\sim 100$ days, and the donor's effective temperature decreases from $\sim 20000\,\rm K$ to $\sim 5000\,\rm K$. At the end of mass transfer, the donor becomes a $\sim 0.7\,M_\odot$ white dwarf.

\begin{figure*}[hbtp]
\centering
\includegraphics[width=0.8\textwidth]{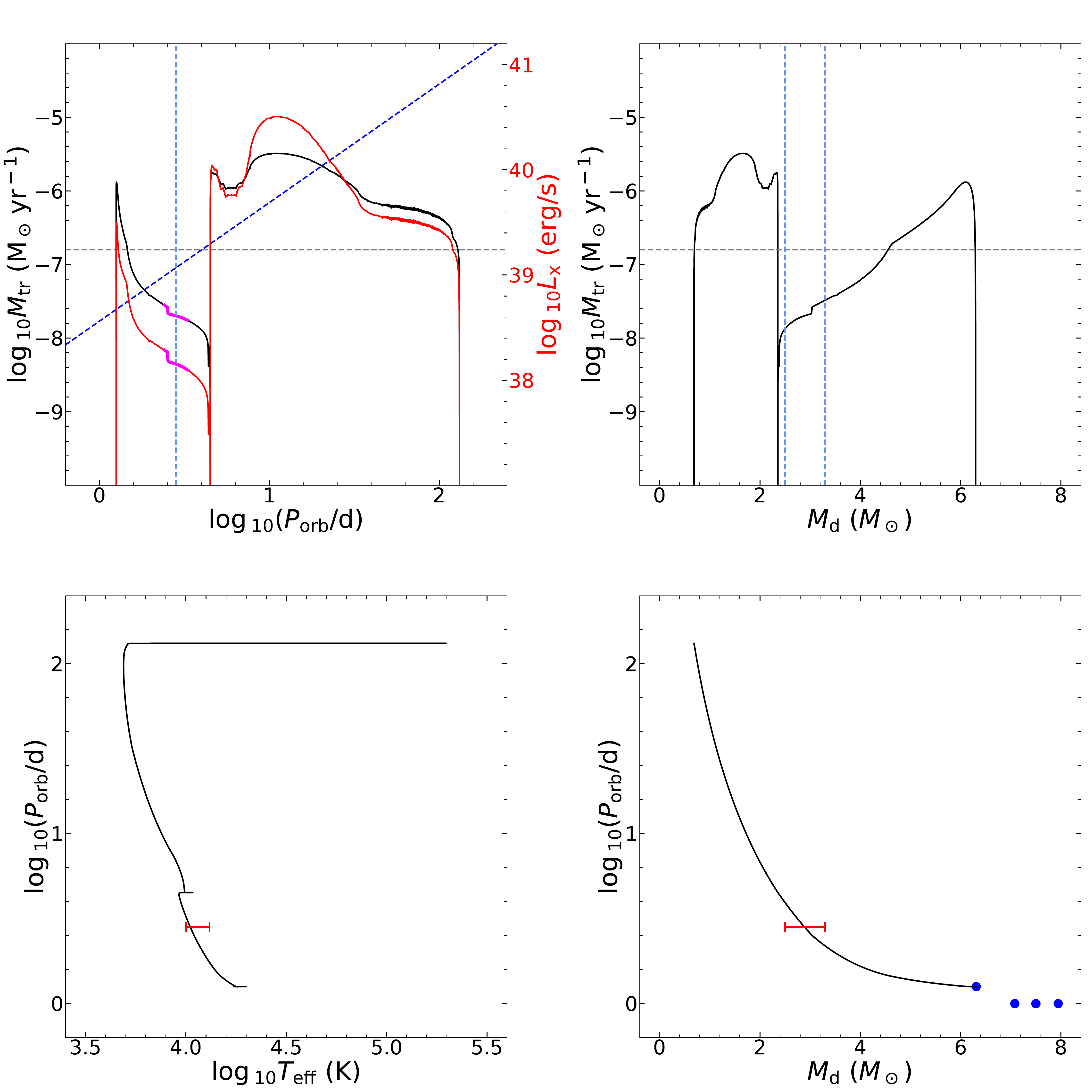}
%\linespread{0.7}
\caption{Similar to Figure \ref{V4641sgrf1}, but for the binary initially containing a $5\,M_\odot$ BH and a $6.3\,M_\odot$ donor in a 1.26-day orbit.
   \label{V4641sgrf2}}
\end{figure*}

\begin{figure*}[hbtp]
\centering
\includegraphics[width=0.8\textwidth]{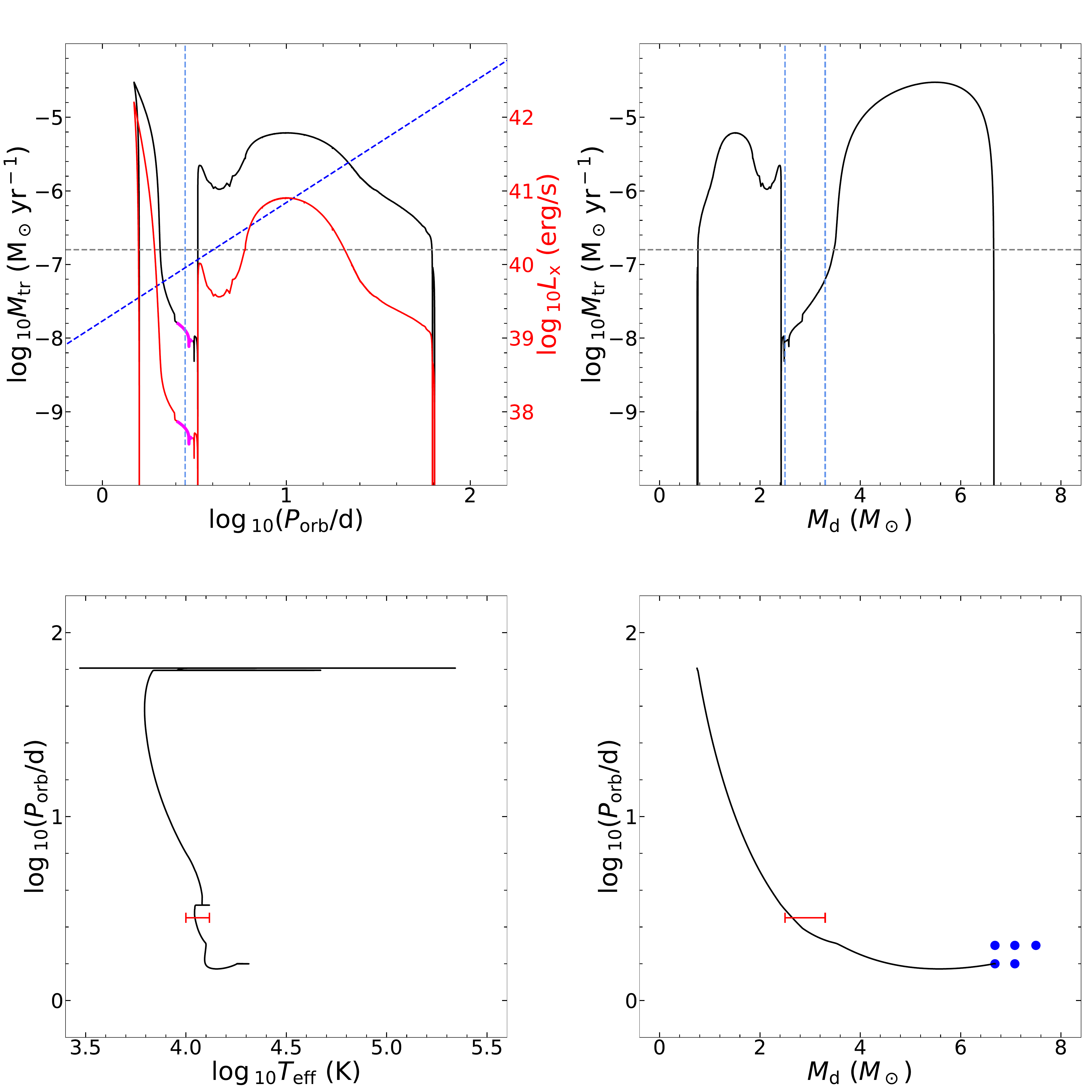}
%\linespread{0.7}
\caption{Similar to Figure \ref{V4641sgrf1}, but for the binary initially containing a $4\,M_\odot$ BH and a $6.7\,M_\odot$ donor in a 1.58-day orbit.
   \label{V4641sgrf3}}
\end{figure*}

\begin{figure*}[hbtp]
\centering
\includegraphics[width=0.8\textwidth]{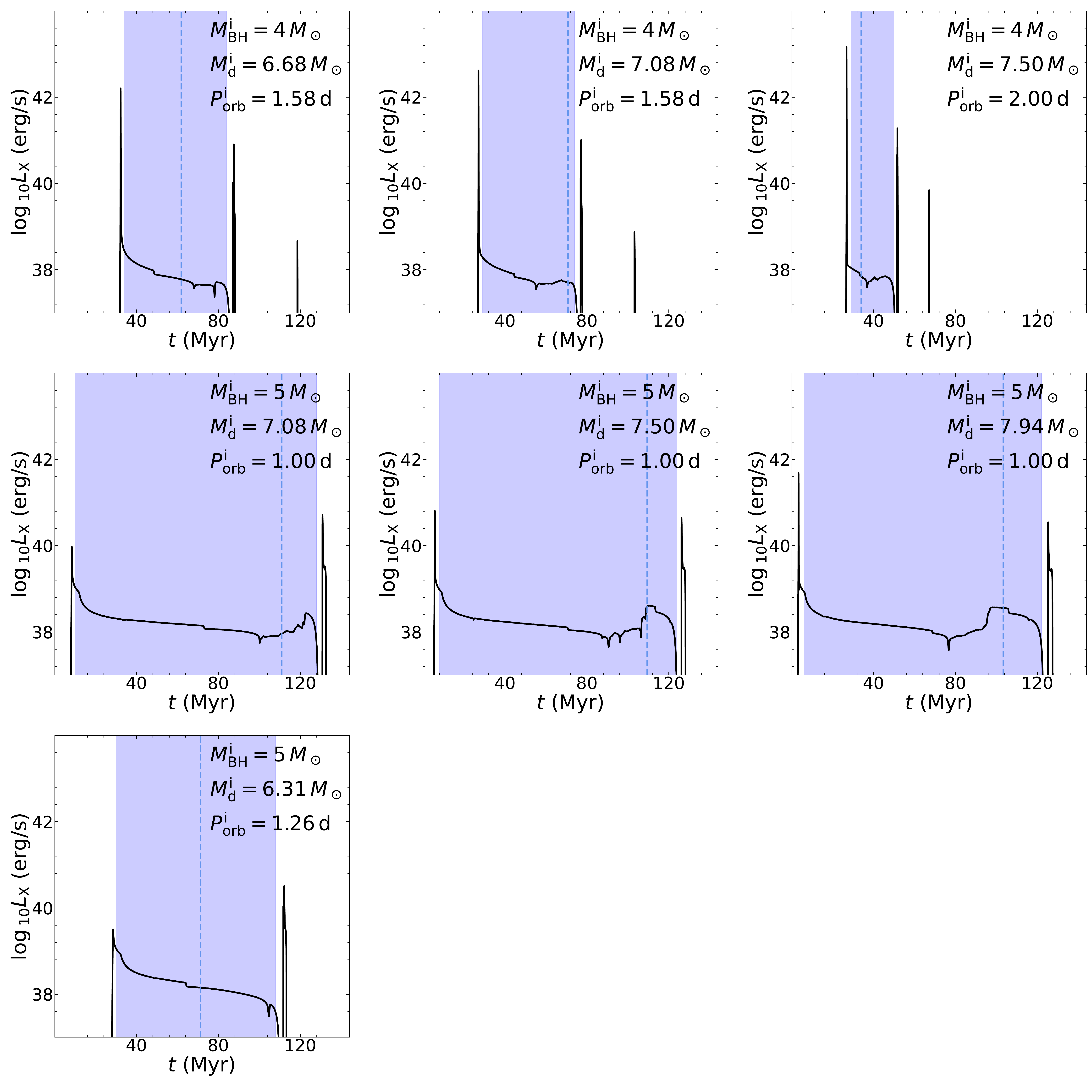}
%\linespread{0.7}
\caption{X-ray luminosity ($L_{\rm X}$) as a function of evolutionary time ($t$) for all binary systems from our library that evolve to match the observed properties of V4641~Sgr. The initial parameters for each track are labeled in the upper-right corner of each panel. The vertical dashed line in each panel marks the evolutionary epoch at which the binary reproduces the current-day observational properties of V4641~Sgr, including donor mass, orbital period, and effective temperature (see also Figures~\ref{V4641sgrf1}-\ref{V4641sgrf3}). The shaded regions highlight the slow mass-transfer phase. In all cases, the matching epoch lies within the slow mass-transfer phase presenting a long-term intrinsic X-ray luminosity of $\sim 10^{38}\rm\,erg\,s^{-1}$.}
\label{V4641sgrf4}
\end{figure*}

Figures~\ref{V4641sgrf2} and \ref{V4641sgrf3} present the evolutionary tracks for two additional binaries with initial parameters of $M_{\rm BH}^{\rm i}=5\,M_\odot$, $M_{\rm d}^{\rm i}=6.3\,M_\odot$, and $P_{\rm orb}^{\rm i}=1.26\,\rm days$, and of $M_{\rm BH}^{\rm i}=4\,M_\odot$, $M_{\rm d}^{\rm i}=6.7\,M_\odot$, and $P_{\rm orb}^{\rm i}=1.58\,\rm days$, respectively. The evolution of these two system differs slightly from that with the higher-mass donor case. Overall, these simulations successfully reproduce the  properties of V4641~Sgr during the slow mass-transfer phase, which lasts over $\sim 10$\,Myr. %The inferred X-ray luminosity reaches $L_{\rm X} \sim10^{38}\rm\,erg\,s^{-1}$ in all these realizations 
Furthermore, for all V4641~Sgr-like systems from our complete evolutionary library (i.e., all systems that can simultaneously reproduce the observed donor mass, orbital period, and effective temperature of V4641~Sgr), the inferred secular X-ray luminosity reaches $L_{\rm X} \sim10^{38}\rm\,erg\,s^{-1}$, far exceeding the observed long-term X-ray luminosity of V4641~Sgr (see Figures~\ref{V4641sgrf4}). 
It should be note that for all matched systems, the simulated mass-transfer rates lie well below the critical rate for disk instability (\citealt{Lasota2001,Lasota2008}; represented by slant dashed lines in the upper-left panel in Figures~\ref{V4641sgrf1}-\ref{V4641sgrf3}). The disks are therefore expected to be thermally and viscously unstable, and hence the X-ray luminosities inferred from the mass-transfer rates should be interpreted as the long-term average rather than as a persistent X-ray emission directly visible to the observer or to the donor. In such a transient state, matter can accumulate in the disk during long quiescent intervals and be accreted during relatively short outbursts. The instantaneous X-ray luminosity and donor irradiation can therefore remain low for most of the time, consistent with the quiescent optical ellipsoidal light curves \citep{MacDonald2014}. Apart from the 1999 event, V4641~Sgr has exhibited lower outburst luminosities than expected. This suggests that the observed X-ray output is unlikely to represent the true long-term accretion power of the system, if the current binary state is indeed described by the matching evolutionary solutions. An obscuration scenario of the system's X-ray emission would naturally explain the discrepancy between the low average X-ray luminosity inferred from observations in past decades and the high secular accretion power suggested by our simulations.

A recent re-analysis of archival VLA imaging suggests that the radio jet position angle is closely aligned with the axis of the extended TeV–PeV emission \citep{Marti2026}, supporting a jet origin of the TeV-PeV emission. The inferred inclination angle of the jet is almost perpendicular to observer's line of sight with $i_j\simeq 89^\circ$. This geometric configuration makes it more plausible that the observed X-ray flux is significantly suppressed relative to the intrinsic accretion luminosity.

Although we do not attempt to derive an obscurer model in this work, previous observations already indicate obscuration at the order-of-magnitude level. In the 2020 outburst, Chandra spectroscopy required partial-covering absorption with hydrogen column density of a few 
times $10^{22}\,\rm cm^{-3}$ and covering fractions of $\sim 0.3–0.4$ \citep{Shaw2022}, while the apparent continuum parameters implied that the intrinsic luminosity had to exceed the observed one by factors of at least 
$\sim 20-40$ to avoid an unphysical inner-disk radius. Our evolutionary calculation extends this picture to long timescales, and the expected intrinsic luminosity is of order $10^{38}\,$erg/s. Compared with an observed long-term average of $<10^{35}\,$erg/s, implying a time-averaged suppression factor of at least $\sim 10^3$. If this interpretation is correct, future observations would again show unusually hot continua for the low apparent X-ray flux, variable reprocessing signatures, and multiwavelength emission that is not simply correlated with the directly observed X-rays.

\section{Implication for the origin of the UHE gamma-ray emission}
\label{sec:discuss}
Previous studies have shown a tight correlation between X-ray luminosity and radio luminosity of BHXRBs in the quiescent state (i.e., $L_{\rm R}\propto L_{\rm X}^q$ with $q\sim 0.5-0.7$) \citep[e.g.,]{Merloni2003, Gallo2003, Corbel2013, Dong2015, Gultekin2019}. Since the radio flux traces the jet kinetic power \citep{Blandford1979, Falcke1995}, this finding implies a strong connection between accretion and jets. Following \citet[][see also \citealt{Fender2004}]{Heinz2005}, we adopt the empirical scaling 
\begin{equation}
L_{\rm jet, k}=1.8L_{\rm Edd}(L_{\rm X}/L_{\rm Edd})^{0.42}(M_{\rm BH}/10M_\odot)^{0.55}.    
\end{equation}
Note that the prefactor 1.8 may be subject to properties of individual BHXRB (such as the jet's opening angle, the proton content, the magnetization, etc) and suffers about an order-of-magnitude uncertainty. Nevertheless, we may estimate the long-term jet power of V4641~Sgr to be $L_{\rm jet, k}\sim 3\times 10^{38}\,$erg/s through this formula. The inferred kinetic luminosity can readily satisfy the required electron kinetic luminosity in the shear acceleration model for the UHE gamma-ray emission \citep{Wan2026}. It may also marginally supply the required PeV proton power in the hadronic model, especially considering that the demanded proton power may be alleviated by a few times in the hybrid model where protons only account for the highest-energy gamma-ray emission \citep{Kleimenov2025}. 
 
On the other hand, extension of V4641~Sgr's gamma-ray spectrum to 800\,TeV suggests efficient particle acceleration operating in the system, boosting energies of particles to at least PeV regime. While the Hillas criterion $E_{\rm H}=eBR\beta_{\rm j}$ gives a necessary condition for particle acceleration, the maximum available particle energy $E_{\rm max}$ may only achieve a fraction of $E_{\rm H}$ in reality. Here $R$ is the size of the acceleration zone and $\beta_{\rm j}$ is the bulk velocity of the jet in unit of the speed of light $c$. Denoting $E_{\rm max}=\eta E_{\rm H}$, the parameter $\eta$ may be understood as the particle acceleration efficiency \citep{Aharonian2002, Wang2025}. Therefore, the condition $E_{\rm max}\geq 1\,$PeV may be translated to a constraint on the magnetic field in the acceleration zone (if the particle acceleration is related to the jet) as $B>1\,{\rm PeV}/\eta eR\beta_{\rm j}$. We may then estimate the requirement of the magnetic luminosity of the jet as $L_{B}=2\pi R^2\beta_{\rm j} c(B^2/8
\pi)>1.6\times 10^{37}(\eta/0.1)^{-2}(\beta_{\rm j}/0.5)^{-1}\,\rm erg/s$. Only a small fraction of the inferred jet's kinetic luminosity being converted to magnetic energy would meet the requirement of PeV  particle acceleration, as long as the particle acceleration efficiency is not too low.

\section{Conclusion}
\label{sec:conclusion}
LHAASO's observation suggests V4641~Sgr as one of the most extreme particle accelerators in our Galaxy and a promising candidate for the long-sought PeV cosmic-ray source. We have performed binary evolution simulations for V4641~Sgr to constrain its long-term mass transfer rate and intrinsic X-ray luminosity. We have found that, within an extensive grid of MESA binary-evolution calculations, all evolutionary tracks consistent with the observed system parameters of V4641~Sgr enter a long-lasting slow mass-transfer phase, with long-term average accretion power and X-ray luminosities of order $10^{38}\,$erg/s. It is significantly higher than the observed average ($L_{\rm X}<10^{35}\,{\rm erg\,s}^{-1}$). Although this does not constitute a model-independent proof of the system’s exact current state, it shows that high intrinsic accretion power is likely a robust outcome of the matching evolutionary solutions.   This discrepancy implies that the X-ray emission from the inner part of the accretion disk is heavily obscured,  an interpretation consistent with previous X-ray and optical studies, likely due to a high-inclination viewing geometry and an extended equatorial outflow or thickened disk.

The inferred intrinsic accretion power, when coupled with empirical correlations between X-ray luminosity and jet kinetic power, yields a jet kinetic luminosity of $L_{\rm jet,k} \sim 3 \times 10^{38}\,{\rm erg/s}$. This comfortably supplies the required energy budget for relativistic electrons ($\sim 10^{37}\,{\rm erg/s}$) in the leptonic shear-acceleration model proposed to explain the extended ultrahigh-energy (UHE) gamma-ray emission. It also marginally accommodates the more demanding proton power needed in purely hadronic scenarios, especially if hybrid models are invoked where protons only contribute to the highest-energy part of the spectrum. Furthermore, the condition for accelerating particles to PeV energies can be satisfied with only a small fraction of the jet kinetic power being converted to magnetic energy, as long as the acceleration efficiency is not extremely low. Thus, the apparent ``energy crisis'' for V4641~Sgr’s UHE gamma-ray emission can be resolved if the system’s true accretion luminosity has been substantially underestimated due to obscuration. Future high-resolution multi-wavelength observations of V4641~Sgr and similar systems, particularly in the X-ray and radio bands, would be crucial to further constrain the obscuring structure and jet properties of V4641~Sgr and similar systems, and provide insights into the nature of this extreme particle accelerator. 

 Finally, we compare V4641 Sgr-like systems with low-mass BHXRBs in terms of their potential roles as UHE gamma-ray emitters and high-energy cosmic-ray accelerators. Our matching solutions indicate that V4641 Sgr descended from a binary with an intermediate-mass donor ($M_d^i\sim 6-8\,M_\odot$). Binary population synthesis simulations \citep{Shao2019} indicate that BH systems with initially intermediate-mass donors ($M_d^i\sim 3-8\,M_\odot$) can have a larger formation rate than those with initially low-mass donors ($M_d^i\lesssim 1-2\,M_\odot$), whereas the latter population constitutes the majority of currently known BHXRBs. This is because the V4641 Sgr-like system of an intermeidate-mass donor is relatively short-lived, so the number of such systems observable at any given time is small. This short evolutionary timescale also implies a high secular mass-transfer rate and hence a high accretion power, allowing V4641~Sgr-like systems to supply energy to jets and energetic particles at a higher rate than typical low-mass BHXRBs. In future work, we will employ binary population synthesis simulations to quantify the relative contributions of different types of BH systems to UHE gamma-ray emission, as well as the overall contribution of BH systems to the Galactic cosmic-ray budgets.

\begin{acknowledgements}
This work is funded by Basic Research Program of Jiangsu under grant No.~BK20250059, National Natural Science Foundation of China under grant No.~12393852, and National Key Research and Development Program of China (grant No. 2023YFA1607902 and 2021YFA0718500). {All input files to reproduce our results are available for download from Zenodo at \dataset[doi.org/10.5281/zenodo.18372798]{https://doi.org/10.5281/zenodo.18372798}.}
\end{acknowledgements}

\bibliographystyle{aasjournalv7}
\bibliography{ms}

\end{document}